# Pushing the Boundaries of Crowd-enabled Databases with Query-driven Schema Expansion

Joachim Selke        Christoph Lofi        Wolf-Tilo Balke

Institut für Informationssysteme
Technische Universität Braunschweig
Braunschweig, Germany

{selke, lofi, balke}@ifis.cs.tu-bs.de

## ABSTRACT

By incorporating human workers into the query execution process crowd-enabled databases facilitate intelligent, social capabilities like completing missing data at query time or performing cognitive operators. But despite all their flexibility, crowd-enabled databases still maintain rigid schemas. In this paper, we extend crowd-enabled databases by flexible query-driven schema expansion, allowing the addition of new attributes to the database at query time. However, the number of crowd-sourced mini-tasks to fill in missing values may often be prohibitively large and the resulting data quality is doubtful. Instead of simple crowd-sourcing to obtain all values individually, we leverage the user-generated data found in the Social Web: By exploiting user ratings we build *perceptual spaces*, i.e., highly-compressed representations of opinions, impressions, and perceptions of large numbers of users. Using few training samples obtained by expert crowd sourcing, we then can extract all missing data automatically from the perceptual space with high quality and at low costs. Extensive experiments show that our approach can boost both performance and quality of crowd-enabled databases, while also providing the flexibility to expand schemas in a query-driven fashion.

## 1. INTRODUCTION

Relational databases have been widely accepted as a mature and potent technology powering numerous applications relying on structured tabular data. But while modern systems show impressive performance and reliability, their strict assumptions regarding data correctness, completeness, and schema adherence hinders their success in many emerging application scenarios. Especially, the recent shift to more social, dynamic, and personalized services on the Web poses a severe challenge for state-of-the-art relational systems. Such services do not only require reliable and efficient retrieval on very large datasets, but also the ability to cope with ambiguous, incomplete, vague, and often uninterpreted data.

Unlike computers, humans are excellent in dealing with perceptual challenges. For instance, we can easily realize that "IBM" and "International Business Machines" refer to the same entity or feel whether a movie is "dramatic." This recently gave birth to *crowd-enabled DBMS* [1–3], which fuse traditional relational technology with the cognitive power of people at query time enabled by crowd-sourcing services like Amazon's Mechanical Turk, CrowdFlower, or SamaSource.

Crowd-enabled databases introduce completely new capabilities: The crowd can be used to complete data missing in tables in a *query-driven fashion*, i.e., a query can be executed despite incomplete data by either filling empty fields or by inserting completely new tuples at runtime. In particular, complex cognitive tasks like reference reconciliation can be performed, and the crowd can even be leveraged to "implement" database operators requiring cognitive abilities like ordering tuples with respect to perceived criteria (e.g., ordering images by visual appeal) or performing perceptual comparisons (e.g., finding an image evoking specific emotions).

But still, the current designs of crowd-enabled databases also show serious drawbacks: Crowd-enabled databases lose many established features of relational systems like roughly predictable execution times, or effective query optimization. Moreover, the actual database schema is still rigid, and will not easily adapt to queries or changing requirements. To fully harness the power of the crowd and to provide flexible and adaptive querying, a certain amount of *schema malleability* [4], [5] is needed. Malleable schemas allow for answering queries using *slightly new attributes*; the required data is automatically obtained during runtime. Up to now, this could only be realized whenever the data for the new column could be derived, relaxed, or merged from already existing columns by exploiting data redundancies (e.g., [6]).

In this paper we show that crowd-enabled databases allow for schema malleability with unpreceded flexibility: Even queries requiring values not covered by any redundancy can still be processed by crowd-sourcing. Unfortunately, naïve implementations of this concept will result in an excessive number of human mini-tasks, potentially resulting in queries with prohibitive performance and unnecessarily high costs. Our work addresses the central challenge of efficient query processing on flexible and malleable schemas in crowd-enabled databases. We introduce an innovative approach that additionally exploits the Social Web instead of relying just on crowd-sourced mini-tasks. Moreover, we argue that the capabilities of this approach go even beyond the vision of malleability, leading to truly *query-driven schema expansion*.

In brief, our approach addresses several open issues and unsolved challenges of current crowd-enabled database designs:

- *Performance:* As outlined by [1], a large number of HITs[1] may have to be issued for each crowd-enabled query. While

---

[1] HIT: Human Intelligence Task, the smallest unit of crowd-sourceable work; many similar HITs are organized in HIT groups.

- HITs are semantically powerful operations, completing large HIT groups may take very long and impose severe performance challenges [7]. We show how a large number of missing values can be obtained while issuing just a few HITs.
- *Scalability*: It has been shown that each requester in a crowd-sourcing platform can only utilize a relatively small human worker pool [1]. This means, HITs cannot be parallelized arbitrarily since the number of simultaneous human computations is capped by the worker pool size. Hence, the scalability of query processing is hampered. Furthermore, if diverse opinions of large numbers of different human workers are required, the worker pool size may not be sufficient anyway. By exploiting also the Social Web, a worker pool larger by several orders of magnitudes is available to the system.
- *Data Quality*: The quality of workers available for crowd-sourcing is hard to control, and thus elaborative quality management is necessary. This usually means executing each HIT several times (e.g., for majority votes), obviously deteriorating query performance. Unfortunately, it is unclear how many times each particular HIT needs to be assigned for a reliable result. We show how our approach can be used during query execution to identify just those tuples which require larger numbers of votes to reach a clear majority judgment, thus saving considerable human effort in all other cases.

## 2. CROWD-DRIVEN SCHEMA EXPANSION

Originally, malleable schemas [4], [5] have been designed to deal with the vague nature of both data and queries in the real world. The rationale is that DBMSs should *try* to answer a query whether the required data is directly available or not—a vision that is deeply shared with crowd-enabled databases. On the data side, providing malleability has been mainly approached using IR techniques mixing structured and unstructured data. On the query side malleability was understood in terms of query relaxation to match more data at runtime [6]. However, both techniques have in common that they can only work if the required missing query attributes closely resemble the available data. This limitation also severely hindered the further success of malleable schemas.

Despite their vision of dealing with incomplete data, current crowd-enabled databases still retain rigid and fixed schemas. They do not live up to their full potential, as filling missing data could be *completely* crowd-sourced: Queries on attributes not yet covered by available data can be answered, too! Then, crowd-enabled databases would support full *crowd-driven schema expansion*.

*Example*: Consider databases containing metadata on movies (like IMDb.com with over two million movie records). Basically, such a database contains *factual* attributes (such as name, year, director, or a list of actors), which are readily available for each movie, but at first may not contain any *perceptual* attributes (i.e., attributes relying on human judgment, such as genre classifications, entertainment value, or excitement rating). For improved service, often human experts are employed to provide this information manually. That means to offer query capabilities like for instance "SELECT name FROM movies WHERE humor $\geq 8$" (i.e., select movies perceived as funny), a judgment of how humorous a movie is has to be crowd-sourced for over two million movies – a truly monumental task. Moreover, each new movie added to the database will require similar HITs.

Fortunately, for most of the popular domains, the crowd has already been busy: literally millions of people voluntarily donate their cognitive abilities by rating, tagging, or reviewing content. This vastly exceeds the work force available to ad-hoc crowd-sourcing services. Therefore, a straightforward idea is to tap into the data available on the Social Web in order to solve the performance problem. Unfortunately, this tremendous volume of human work can hardly be used directly in a crowd-enabled database.

The innovative idea of this paper is that the sum of all available user-provided metadata of content can be transformed into *perceptual spaces*; a space encoding all users' opinions and perceptions with respect to certain items in some aggregated form (this concept is explained in detail in the next section). Unfortunately, also this perceptual space cannot be used directly: The individual perceptual attributes must be extracted from the space in an approximate and semi-automatic fashion. However, once this extraction has been performed, a wide variety of perceptual attributes can be directly derived from the space, remedying the apparent performance issues of crowd-sourcing.

Our approach is not limited to the movie domain used in the example. Nearly all perceptual attributes can be supported as long as they influence the rating behavior of users. As a consequence, purely factual information (e.g., missing email addresses) cannot be derived from the numerical ratings provided by users and needs to be crowd-sourced individually.

Overall, our contributions in this paper can be summarized as:

- We extend crowd-enabled databases by the concept of crowd-driven schema expansions. Using the concept of perceptual spaces, we show how they can actually be built from the Social Web in a flexible and meaningful way.
- Once built, we show how perceptual attributes can be dynamically extracted from this space in a very efficient manner. In particular, large numbers of values required for high-quality schema expansion can be derived by crowd-sourcing the training phase of the classification algorithm. Here, our aim is not to propose a novel learning technique but rather to point out the benefit of data collected from the Social Web for improving crowd-enabled databases.
- We outline how our approach can also benefit other open issues of crowd-enabled databases (i.e., performance, scalability, and data quality issues).

## 3. THE CASE FOR PERCEPTUAL SPACES

Next we present the concept of perceptual spaces derived from the Social Web and provide the necessary mathematical foundations.

### 3.1 The Social Web and the Crowd

The Social Web is on a steep rise. Originally developed for exchange between like-minded people, Social Web platforms have become a major innovator of technology. With new services being established continuously, and many older ones growing in popularity, a significant shift in user behavior has occurred: People got accustomed to an active and contributive usage of the Web. Users feel the need to express themselves and connect with like-minded peers. As a result, social networking sites like Facebook amassed over 800 million users[2]. At the same time, there are countless special-interest sites for music, movies, art, or anything that is of interest to any larger group. But the real revolution lies in the way people interact with these sites: Following their social nature, millions of people discuss, rate, tag, or vote content and items they encounter on the Web. Therefore, "I Like" buttons, star scales, or comment boxes are omnipresent on today's Web.

---
[2] http://www.facebook.com/press/info.php?statistics

This development can be seen as a massive crowd-sourcing task gathering perceptual data from its workers. In contrast to explicit crowd-sourcing as used by crowd-enabled DBMS, generating data in the Social Web follows different rules: People in the Social Web are entirely intrinsically motivated, and contribute voluntarily. For example, a user finding an interesting online news article might vote for that article on his preferred social site, while a user leaving the cinema after a particular bad movie experience may log onto a movie database, rating the movie lowly, and venting his disappointment in a short comment or review. Therefore, the biggest hindrance in directly using the Social Web as a reliable source of data is that user contributions can neither be controlled nor do they follow a strict schema or guideline. Thus, with respect to processing this data automatically, most of this vast wealth of valuable information just lies dormant.

In this paper, we will focus on exploiting user-provided feedback in form of *rating data*, i.e., feedback of users judging certain items on a numerical scale. Rating data has several significant advantages on which we can capitalize: (1) Rating data is easy to handle algorithmically and can be exploited by relying on proven assumptions from the area of recommender systems (see next section). (2) Rating information is ubiquitous in the Social Web due to its simplicity and low required effort from a user's perspective. (3) As we will demonstrate, rating data contains rich and valuable information. In the experimental section of this work, we will perform detailed experiments on movie ratings, but we will also showcase our algorithms in the domains of restaurants and board games. Furthermore, we will also discuss the availability of user rating within the Social Web.

### 3.2 Semantics of Perceptual Spaces

Perceptual spaces are built on the basic assumption that each user within the Social Web has certain personal interests, likes, and dislikes, which steer and influence his/her rating behavior [8]. For example, with respect to movies, a given user might have a bias towards furious action scenes; therefore, he/she will see movies featuring good action in a slightly more positive light than the average user who doesn't care for action. The sum of all these likes and dislikes will lead to the user's overall perception of that movie, and will ultimately determine how much he enjoyed the movie and therefore, will also determine how he rates it on some social movie site. Moreover, the rating will share this bias with other action movies in a systematic way. Therefore, one can claim that a perceptual space captures the "essence" of all user feedback, and represents the shared as well as individual views of all users. A similar reasoning is also successfully used by recommender systems [9], [10].

Now, the challenge of perceptual spaces is to reverse this process: For each item which was rated, commented, or discussed by a large number of users, we approximate the actual characteristics (i.e., the systematic bias) which led to each user's opinion.

Formally, we implement this challenge by assuming that a perceptional space is a *d*-dimensional coordinate space satisfying the following constraints: Each user and each item is represented as a point in this space. The coordinates of a user represent his personality, i.e., the degree by which he likes or dislikes certain characteristics. The coordinates of items, in contrast, represent the profile of that item with respect to same characteristics. Items which are perceived similar in some aspect have somewhat similar coordinates, and items which are perceived dissimilar have dissimilar coordinates.

Next, we assume that a user's overall perception of an item is anti-proportional to the distance of the user and item coordinates, i.e. the "best movie of all times" from a given user's perspective has the same coordinates as the user him-/herself. Of course, a user's likes and dislikes may be slightly unstable due to moods; but on average, this assumption is stable enough. Thus, the characteristics which are of interest with respect to a certain query posed to a crowd-enabled database (such as a movie's genre, the quality of action sequences, etc.) generally correspond to some (or a set of) dimensions of the perceptual space, but unfortunately must still be extracted from it by learning the semantic of the desired attribute from the crowd (see Figure 1 for a simplified example space featuring movies).

This leaves two big challenges, which we address next:

- How can the feedback of users in the Social Web be used to construct a perceptual space with the described properties?
- How can the characteristics of an item be extracted given its coordinates in the perceptual space?

### 3.3 Building Perceptual Spaces

As argued above, in the following, we will focus on perceptual spaces constructed from rating data. However, the general approach presented below can be adapted to build perceptual spaces from text collections (e.g., product reviews), semi-structured annotations (e.g., tags), or other data sources of the Social Web. The challenge to be addressed in this section thus can be summarized as follows: Given a collection of ratings, how to construct a meaningful perceptual space from it?

Our approach heavily relies on factor models, a powerful technique that recently received a lot of attention in recommender systems research [10]. Factor models have originally been developed to estimate the value of non-observed ratings for the purpose of recommending new (yet unrated) items to existing users. However, some initial studies we conducted in [11] indicated the potential benefit of factor models beyond recommendation tasks.

Basically, the factor model approach assumes that users and items each can be represented by a *d*-dimensional real vector and each rating is a function of the corresponding user vector and item vector. This assumption reflects established models of human preference and is well-accepted in recommender systems research [10]. All model parameters are estimated by minimizing a cost function measuring the deviation between the actual observed ratings and those predicted by the model. By stating the above as an optimization problem and solving it, user and item vectors can be found that fit the given rating data best.

Formally, given a large number of ratings, where each rating is a triple $\langle movie\_id, user\_id, score \rangle$,[3] we assume that there are $n_M$ movies (a single movie is denoted by $m$), $n_U$ users (a single user is denoted by $u$), and ratings $r_{m,u} \in \mathbb{R}_\square := (\mathbb{R} \cup \{\square\})$.[4] Typically, the total number of actually provided ratings is very small compared to the number of possible ratings $n_M \cdot n_U$, usually lying in the range of 1–2%. For representing all movies and users as coordinates in $\mathbb{R}^d$, we need to find two matrices $A = (a_{m,k}) \in \mathbb{R}^{n_M \times d}$ and $B = (b_{u,k}) \in \mathbb{R}^{n_U \times d}$ such that $a_m = (a_{m,1}, \dots, a_{m,d})$ are the coordinates of movie $m$ in the perceptual space, and $b_u = (b_{u,1}, \dots, b_{u,d})$ are those of user $u$.

---

[3] For illustration, but without loss of generality we focus on movies.

[4] The symbol □ indicates that the user did not rate the specific item yet.

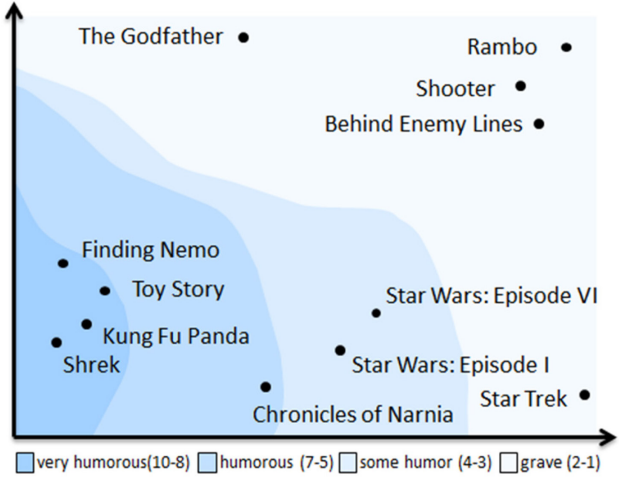

Figure 1: Example perceptual space
A simplified perceptual space in $\mathbb{R}^2$. While the dimensions do not convey any semantics directly, a judgment of a movie's humor can be extracted.

Moreover, factor models assume that there is a rating function $f : \mathbb{R}^d \times \mathbb{R}^d \to \mathbb{R}$ such that $r_{m,u} = f(a_m, b_u) + \epsilon_{m,u}$, for any movie $m$ and user $u$ with $r_{m,u} \neq \square$, where $\epsilon_{m,u}$ is a random noise term modeling external rating deviations not captured by the model. The model's prediction of user $u$'s rating of movie $m$ then is $\hat{r}_{m,u} \coloneqq f(a_m, b_u)$. To measure the fit between the available rating data and the model, a cost measure $C : \mathbb{R}^{n_M \times d} \times \mathbb{R}^{n_U \times d} \times \mathbb{R}_\square^{n_M \times n_U} \to \mathbb{R}$ is introduced that usually takes the form $C(A, B, R) = \sum_{m,u \text{ s.t. } r_{m,u} \neq \square} c(r_{m,u}, \hat{r}_{m,u})$, where $c : \mathbb{R} \times \mathbb{R} \to \mathbb{R}$ measures the individual deviation wrt. the rating pair $(m, u)$.

Probably the most elementary factor model is the SVD model, which is mainly used for collaborative filtering. It defines $f$ as the scalar product on $\mathbb{R}^d$ and sets $\epsilon_{m,u} = \epsilon$ as a zero-mean Gaussian that is independent of $m$ und $u$. As cost function the very popular mean squared error (MSE) is used, i.e., we set $c(x, y) = (x - y)^2$. As our goal is to minimize the costs, this approach directly leads to the following optimization problem:

Minimize (in $A, B$) $\sum_{\substack{m,u \text{ s.t.} \\ r_{m,u} \neq \square}} (r_{m,u} - a_m \cdot b_u)^2 + \lambda \cdot (a_m^2 + b_u^2),$

where $\lambda \geq 0$ is a regularization constant that is determined by the variance of the Gaussian error term. In practice, the dimensionality $d$ and the regularization parameter $\lambda$ are determined by means of cross-validation, where $d$ typically falls into the range between 50 and 200. While the SVD model proved to be highly effective for collaborative filtering tasks, it is not clear how a meaningful similarity measure on movie pairs could be derived from it.

Instead, we propose a modified version of Euclidean Embedding presented in [12], which is designed around the standard Euclidean distance. As discussed in the previous section, we want the distance between movie $a_m$ and user $b_u$ to be small if user $u$ likes movie $m$; otherwise, it should be large. To account for general effects independent of personal preferences, for each movie $m$ and user $u$, we introduce the model parameters $\delta_m$ and $\delta_u$, which represent a generic rating bias relative to the average rating $\mu$.

For example, assume that the average rating across all users and movies is $\mu = 6.2$ (given a rating scale from 1 to 10). Then, a better-than-average movie $m$ with a movie specific rating average of 8.4 has a positive bias $\delta_m = 2.2$. Furthermore, a critical user $u$ who generally provides rather more negative ratings than his peers (regardless of the movie) will show a negative user bias, say, $\delta_u = -1.6$. Assuming that this user is absolutely neutral with respect to the actual characteristics of the movie, his estimated rating will be $r_{m,u} = 6.8$. Any aberration from this estimated value can be attributed to the user not being fully neutral, i.e., the movie's characteristics are agreeing or clashing with some of the user's likes or dislikes (this is represented by the distance between the coordinates of the movie and user in the perceptual space).

These observations lead to the following rating estimation term:

$$\hat{r}_{m,u} = \mu + \delta_m + \delta_u - d_E^2(a_m, b_u),$$

where $d_E(\cdot, \cdot)$ denotes the Euclidean distance in $\mathbb{R}^d$. Here, a possible drawback of relying on the Euclidean distance becomes apparent: Ratings can only negatively be influenced by the movie–user distance. However, our experiments in Section 4 show that this simplification still leads to high quality results.

For the sake of algorithmic simplicity, we propose to use the same random noise term and cost measure as in the simple SVD model (zero-mean Gaussian and MSE), which leads to the following least squares optimization problem:

Minimize (in $A, B, \delta$)

$$\sum_{\substack{m,u \text{ s.t.} \\ r_{m,u} \neq \square}} \left(r_{m,u} - [\mu + \delta_m + \delta_u - d_E^2(a_m, b_u)]\right)^2 + \lambda \cdot (d_E^4(a_m, b_u) + \delta_m^2 + \delta_u^2).$$

The second term of this minimization problem is used for regularization in order to avoid overfitting. Here, we soften the impact of extreme biases and very large distances. This optimization problem can be solved efficiently using stochastic gradient descent or alternating least squares methods, even on large data sets [13]. On each data set we investigated, we determined the parameters $d$ and $\lambda$ by means of cross-validation on the rating data only. However, we found the exact choice of $\lambda$ to be of minor importance in our experiments ($\lambda = 0.02$ worked well across many different data sets). Similarly, the specific choice of $d$ does not significantly influence the properties of the space as long as $d$ is large enough. We found $d = 100$ to be a good choice for most data sets, and use it in the following.

### 3.4 Expanding Schemas

Our major contribution is to allow flexible and expandable schemas in crowd-enabled databases. For realizing this goal, the only remaining challenge is to extract the correct data from a perceptual space computed according to the previous section. Unfortunately, the data is hidden in the space, and therefore this task can neither be performed directly nor fully automatically:

Assuming a query involving a yet-unknown perceptual attribute (e.g. in the movies domain: humor, suspense, imaginativeness, etc.), we first have to understand what this new attribute means. This is best implemented by providing a "gold sample"; i.e. for a small set of movies, the correct judgment of the desired attribute is provided by human experts. This task can easily be crowd-sourced using the default capabilities of a crowd-enabled DBMS. However, ensuring a high quality of the sample is important. Therefore, trusted workers (i.e., workers who have proven their honesty and knowledge) should be used. Moreover, result quality should be controlled using majority votes or a similar technique.

*Example (cont.)*: To answer our previous query "find the most humorous movies", in our schema expansion system, missing judgments are not crowd-sourced directly. Instead, we rely on the perceptual space provided by numeric user ratings as provided by IMDb.com (users may rate each movie on a scale between 1 and 10 stars, expressing if they liked the movie or not; usually, several thousand users rate each movie). The extraction process for judging humor must be manually trained before it can be applied to the perceptual space, thus a small training set is crowd-sourced: Crowd workers have to provide reliable judgments for, say 100 movies. Then, a numeric judgment for humor can be extracted from the perceptual space for all two million movies without additional user interaction, and with very high accuracy.

From here on, the remainder of the extraction process can be performed automatically: The gold sample can be used to train a classifier or regression machine, which in turn, will be able to approximate the missing values of all other tuples of the database quickly. Generally when extracting numeric judgments from a perceptual space, we suggest to use Support Vector Regression Machines (SVMs) [14], which are a highly flexible technique to perform non-linear regression and classification, and have been proven to be effective when dealing with perceptual data [15]. After training this machine learning algorithm with our crowd-sourced gold sample, based on the perceptual space the algorithm establishes a non-linear regression function. This regression function will finally provide all missing data required for the schema expansion. A graphical overview of the whole workflow is shown in Figure 2: In contrast, basic crowd-enabled databases can only rely on the crowd-sourcing service to provide the missing data.

## 4. EVALUATION

We performed an extensive evaluation of our proposed approach for crowd-driven schema expansion focusing on three aspects:

- Quality and costs of crowd-driven schema expansions relying on crowd-sourcing missing data directly (as a baseline).
- Quality and costs of crowd-driven schema expansion using our method of extracting data from perceptual spaces.
- Quality improvements achieved by cleaning directly crowd-sourced data with our perceptual spaces approach.

**Data Set and Scenario**

To measure the accuracy of the perceptual data elicited during our schema expansion experiments by different approaches, we need a large high-quality reference dataset actually featuring perceptual attributes. Furthermore, we require an extensive collection of user ratings for building perceptual spaces. Therefore, we continue using the movie domain as running example: Here, the community spent considerable efforts for manually providing perceptual judgments and also extensive user ratings have been published.

For building our reference data and to establish ground truth, we rely on the three leading movie databases: Internet Movie Database (IMDb), Netflix, and Rotten Tomatoes (RT). While all three data providers actively employ experts for annotating movies with perceptual attributes, none of them actually spent the monumental effort to elicit fine-grained numeric attributes, like those provided by our approach (e.g., judgments of a movie's humor, action content, or character depth). Unfortunately, the database Clerkdogs, which tried to build such a data set manually, is no longer available for reference. Due to the enormous number of movies and required expert knowledge, we also abstained from the idea of annotating the data by ourselves (see evaluations in Section 4.1).

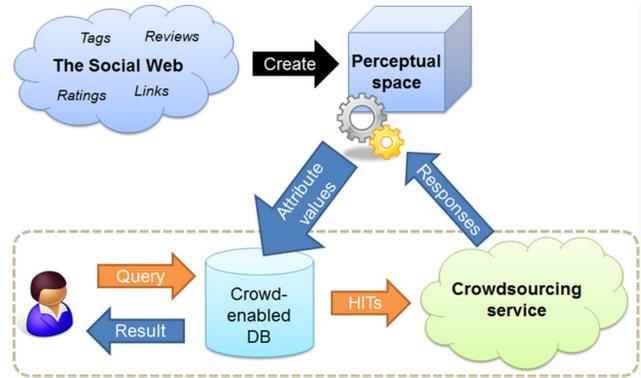

**Figure 2: Overview of crowd-driven schema expansion**

Therefore, we rely on genre classifications (e.g., comedy, action movie, thriller, science fiction), each movie may belong to multiple genres). Genres classifications are a popular and well-accepted classification scheme for movies focusing on those high-level properties that are of major relevance to customers [16]. Thus, they summarize the general perception of a given title. But classifying movies by genre is a challenging task, and neither did the three movie databases provide genre classifications for all their movies, nor did they agree all the time in their judgment. Thus, our reference dataset is comprised of all those movies which are a) contained in all three databases, and b) have a genre classification in all three databases. The actual "correct" genre classification was derived using majority votes on the three expert sources.

For actually building conceptual spaces as described in Section 3.3, we need a large number of real world user ratings for the movies in the reference set. We rely on the rating data published during the Netflix Prize competition in 2006, containing a collection of 103,297,638 ratings (on a five-point scale) of 17,770 movies provided by 480,189 Netflix customers. Overall, we identified 10,562 titles that are covered by the Netflix Prize data set and also have a genre classification in all three expert movie databases, resulting in a total number of 85,651,367 matching ratings.

As a scenario for crowd-driven schema expansion, we built a table containing the factual part of the reference data set (e.g., movie_id, name, year, etc.). The query triggering the schema expansion used throughout our evaluation is to filter all comedy movies; but no genre classification is given in the experiments upfront. Therefore, the schema expansion introduces a new perceptual binary attribute *is_comedy*: "SELECT * FROM movies WHERE is_comedy = true". Using the expert genre classification from the reference dataset, we now can reliably measure the accuracy of different crowd-based approaches for obtaining the missing data.

### 4.1 Quality Issues of the Crowd

In the first set of evaluations, we focus on the quality of naïvely crowd-sourced schema expansions. Performance and cost aspects are discussed in Section 4.2 in comparison to perceptual spaces.

Our evaluation scenario is as follows: In order to constrain the costs of crowdsourcing, we randomly selected 1,000 movies of the 10,562 movie in our reference data set (the same sample is used in all subsequent experiments). For these 1,000 movies, we try to answer the example query "SELECT * FROM movies WHERE is_comedy = true"; i.e., the attribute *is_comedy* needs to be added to the database, thus its values have be crowd-sourced for all 1,000 tuples. This results in multiple HITs, each containing 10 movies which are to be classified by the workers. In order to reliably measure worker' accuracy and consistency, as well as estimating the overall data quality of this approach, each single movie is classified by 10 different workers. Thus, in total of 10,000 judgments need to be obtained.

We performed three experiments, each dispatching HITs using the Crowdflower crowd-sourcing service and executed on Amazon's Mechanical Turk infrastructure. Please note that we did not rigorously control environmental variable like time of day, or day of week. While this may lead to changing worker populations, our experiments still clearly show the issues of various techniques.

*Experiment 1)*: Each HIT contained 10 movies to be classified. For each movie, three options could be selected by the workers (no default option): "Movie is a comedy", "Movie is not a comedy", and "I do not know this movie".

We instructed the workers to only classify movies they really knew (furthermore, we made clear that workers are paid in any case, even if they do not know the movie). Moreover, we asked to provide a personal judgment, i.e., if a worker knew the movie, he/she was supposed to classify it without using additional resources (especially, without looking the movie up on the Web).

For each HIT we paid $0.02. Thus, the overall costs for obtaining 10 judgments per movie are $20 (plus a small service fee paid to Crowdflower). Please note that the movies in this evaluation were a true random sample from our reference data set, i.e., besides some well-known movies like "Superman (1979)" or "Rocky (1976)", the sample also contains many less famous movies. By performing a quick survey on our students, we estimated that an average person will usually know 10–20% of the presented movies. Furthermore, we know from our reference data that 30.1% of the 1,000 movies are comedies according to experts.

Unfortunately, the results of this first evaluation are quite disillusioning with respect to the crowd's quality and reliability: At first glance, one could observe that 62% of the 10,000 movie judgments were "This movie is a comedy", 24% were "No comedy", and only 14% of all judgments admitted that they did not know the movie. Most likely, many workers did not perform this task honestly, and simply selected one of the first two answer options.

This results in very poor overall quality of the collected data: For reaching the final crowd classification, a majority vote ignoring "don't know" answers is performed. Movies are regarded as being not classified at all, whenever no single judgment is available, or no majority could be found due to a tie. This is the case for 107 movies, i.e., 893 have been classified by the crowd, of which 533 have been classified correctly with respect to the reference data. This is 59.7% of all classified movies—a fairly poor result. This task was completed in 105 minutes, i.e., on average, 95 judgments have been provided per minute by 89 different workers.

This observation closely matches surveys in psychology [17]: Our data quality problem can be attributed to the perceptual nature of our HITs, as answers cannot be controlled rigorously: a subjective judgment is asked which is neither right nor wrong by itself. Furthermore, we have to accept that users may not know a presented item and hence cannot judge it. unfortunately, this is also recognized by many crowd workers, motivating them to abuse the system to increase their income. In contrast, when dealing with factual data, "gold questions" can be used where the correct answer is known upfront. These gold questions can be used to penalize dishonest workers, and thus increase the data quality.

The effect of dishonest workers can also be affirmed when analyzing the judgments of individual workers. Two major groups of users are clearly visible: a) users which supposedly knew nearly every movie (94%), no matter how obscure, and judged them as being comedies in 56% of all cases, and b) users who knew only roughly 26% of all movies (which is more realistic); and of those movies they actually knew, judged 32% as comedy and 68% as not comedy, reflecting the real ratio found in the reference data.

Obviously, the second group of workers honestly performed our HITs, while the first group just abused the system. We conclude that significantly greater care is required when dealing with perceptual crowdsourcing tasks compared to factual ones.

*Experiment 2)*: In our second experiment, we adjusted the setting to increase quality. While the general constraints of the HITs remained similar (task, options, payment, etc.), we tried to exclude all unreliable workers. When analyzing the first experiment, we grouped the unreliable users by country. We observed that nearly all malicious workers originated from only very few countries. In this experiment, we excluded those countries explicitly by adapting our HIT definition accordingly. Although this heuristic seems very crude, it resulted in drastically increased accuracy.

Majority votes with respect to "is a comedy" resulted in 636 movies classified correctly. However, 199 movies could not be classified at all. Of all movies that could be classified, 79.4% have been labeled correctly. Furthermore, only few workers claimed to know more than 30% of all movies, hence we assume that most workers can trusted (which does not imply that they provide correct classifications as this is still a challenging cognitive task). This task was completed in 116 minutes by 27 different workers.

Although excluding unreliable users drastically improves accuracy, a new problem becomes obvious: Data can only be elicited if workers can perform an informed judgment for the given item. This is not a problem for typical perceptual crowd-sourcing tasks often found in literature (e.g., classifying facial expressions or ranking images), but becomes indeed an issue when dealing with expandable database schemas. Here, often expert knowledge is required, which is hard to find in the crowd worker population. Therefore, reliable perceptual judgments for *all* items in a database are hard to obtain by means of crowdsourcing.

*Experiment 3)*: Next, we turned the genre classification into a factual crowd-sourcing problem: We again allowed untrusted workers but removed the "don't know" option and instructing workers to look up the correct genre on the Web—most likely in one of the big Internet movie databases, which we also used to create our reference data. Thus, high accuracy can be expected. As this task is more time-consuming, we paid $0.03 per HIT.

The goal of this evaluation is to demonstrate the effects of additional control possible when using factual data. By adding 100 "gold questions", we can identify unreliable workers during query execution time: 1,100 movies had to be classified but for 100 of them (the gold questions), the correct classification was known to the crowd-sourcing service. This 10% ratio of gold-to-tasks matches the generally recommended best-practices[5]. Workers could not distinguish gold questions from normal tasks. Furthermore, workers who repeatedly classified the gold questions incorrectly were automatically excluded.

---

[5] See http://crowdflower.com/docs/gold

**Table 1. Classification accuracy for direct crowd-sourcing**

The number of classified movies (i.e., majority found; out of 1,000), the percentage of correctly classified movies (out of all classified movies), and the total time needed for generating 10,000 judgments.

| Evaluation | #Classified | %Correct | Time |
|---|---|---|---|
| Exp. 1: All | 893 | 59.7% | 105 min |
| Exp. 2: Trusted | 801 | 79.4% | 116 min |
| Exp. 3: Lookup | 966 | 93.5% | 562 min |

**Table 2. Three example movies and their five nearest neighbors in perceptual space**

| **Rocky (1976)** | **Dirty Dancing (1987)** | **The Birds (1963)** |
|---|---|---|
| Rocky II (1979) | Pretty Woman (1990) | Psycho (1960) |
| Rocky III (1982) | Footloose (1984) | Vertigo (1958) |
| Hoosiers (1986) | Grease (1978) | Rear Window (1954) |
| The Natural (1984) | Ghost (1990) | North By NW (1959) |
| The Karate Kid (1984) | Flashdance (1983) | Dial M f. Murder (1954) |

The results of this evaluation are as follows: 903 movies have been classified correctly, 34 movies could not be classified at all. Of the 903 available classifications, we found 93.5% to be correct. Furthermore, this experiment took notably longer and could only be finished after 562 minutes (51 workers). This slow speed can be attributed to the fact that each HIT demanded considerably more effort from the users, increasing the time per HIT, and also lowering the attractiveness of the overall job.

*Summary*: The most important metrics of the last three evaluations are shown in Table 1. We demonstrated that worker quality needs to be controlled. If a large number of malicious users accept the HITs, result quality strongly declines despite majority votes (Exp. 1). If strict control is possible (i.e., there is a large number of known "correct" answers), gold questions can be used to increase quality (Exp. 3). However, this kind of control is usually not possible for perceptual attributes. Therefore, relying on the judgments of honest and trusted workers as in Exp. 2 can be considered as a realistic baseline for crowd-driven schema expansions, resulting in 636 out 1000 movies classified correctly after 10,000 judgments, at a cost of $20.

## 4.2 Boosting the Crowd

In our second set of evaluations, we investigate the benefit of perceptual spaces for crowd-enabled databases. The perceptual space with 100 dimensions used for this purpose is built according to Section 3.3, and is based on user ratings from the Netflix Prize challenge (103M ratings given by 480k users on 17k movies). Interestingly, a straightforward implementation of our approach on this rather large dataset only needed about 2 hours to build the perceptual space on a simple notebook computer. For time-critical applications parallelization techniques are quite easy to exploit.

**The Perceptual Space**

For a better understanding of the coordinates and semantics in perceptual spaces, let us first consider the example in Table 2: Here, three popular movies are shown, along with their respective five nearest neighbors (with respect to Euclidean distance) in our perceptual space. The first column shows movies similar to the famous sports drama Rocky. And indeed, all these movies feature underdog protagonists involved in sports. The second group of movies contains well-known formulaic romantic dramas, which are indeed similar in style to Dirty Dancing. The third group of movies contains classical psychological thrillers directed by Alfred Hitchcock. While the third list possibly could have been generated from metadata only (e.g., year of release and director), deriving the other two definitely requires a deep understanding of a movie's perceptual features. This example already indicates that essential parts of human perception are encoded into the space.

We did not perform any data cleaning or spam detection on the rating data. However, as the Social Web is susceptible to spamming and data quality issues, our approach might benefit from implementing existing anti-spam techniques such as [18]. Additionally, most Social Web services take care of spam detection and banning malicious users, which makes our task easier.

To verify these initial results, we performed several user studies where we directly measured perceived similarity by judging pairs of movies. These studies are beyond the scope of this work and will be published elsewhere. In short, we can conclude that distances measured in the perceotual space correlate to the general user consensus by a Pearson correlation of 0.52. This is quite a large number as each individual user showed just an average Pearson correlation with the consensus of 0.55. Thus, distances measured in the perceptual space are as accurate as the perceived similarity of many users. In essence, we have learned that the perceptual space does not perfectly resemble human perception, but still captures major properties of the domain at hand.

**Perceptual Spaces for Boosting Crowd-Sourcing**

Next, we demonstrate that perceptual spaces significantly boost the performance of direct crowdsourcing. In particular, we show how to increase coverage (i.e., the number of classified movies) and accuracy at dramatically reduced costs and runtime.

Here, as the *is_comedy* attribute of our sample query holds only Boolean values, a slightly simpler algorithmic implementation of the extractor described in Section 3.4 is possible. Instead of relying on non-linear regression, we can use an SVM classifier [19]. We found a non-linear Radial Basis Function kernel to be useful. In order to train the SVM, we have to rely on training samples with existing judgments for the value of the new schema attribute *is_comedy*. In case of our example query, we just need a small set of comedy movies and a small set of non-comedy movies. Generating such training data can be crowd-sourced at query time similar to the experiments in the last subsection.

We built upon Experiments 1–3 and incrementally used the classification results as training data for deriving genre classifications from our perceptual space. During the crowd-sourcing process, we periodically used the movies currently showing a clear majority to train the SVM. After that, we classified all 1,000 movies (even those that none of the crowd workers knew) and measured the accuracy. For a comprehensive overview of all results of the next three experiments over time, please refer to Figure 3. Due to highly varying runtimes of the three reference experiments, we chose time axes relative to the total runtimes. For all results over money spent, please refer to Figure 4.

*Experiment 4)*: This experiment is based on the judgments obtained in Experiment 1 (i.e., including judgments from a large number of untrustworthy workers). We start with an empty training set, and every 5 minutes, all movies currently classified by the crowd-workers are added to it. Thus, the training set grows over time. At the end of the experiment after 105 minutes, the training set contains all 893 movies classified during Experiment 1. Furthermore, after expanding the training set, we re-train the SVM classifier (this takes roughly 0.5 seconds on a standard notebook computer). After every training step we classify all 1,000 movies using the model learned by the SVM. The obtained genre assignments are the final output of our method. This way, we are able to fix labeling errors in the training data.

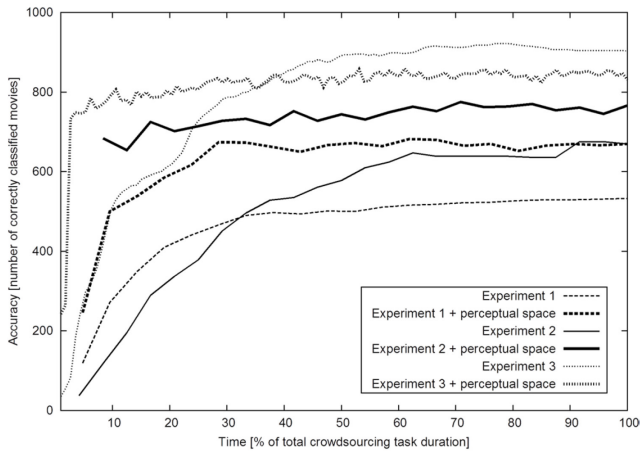

**Figure 3: Correctly classified movies over time**
When using perceptual spaces, a good classification is available after short time. Please that the time axis is relative to the overall runtime (experiment 1: 105 min, experiment 2: 115 min, experiment 3: 562 min).

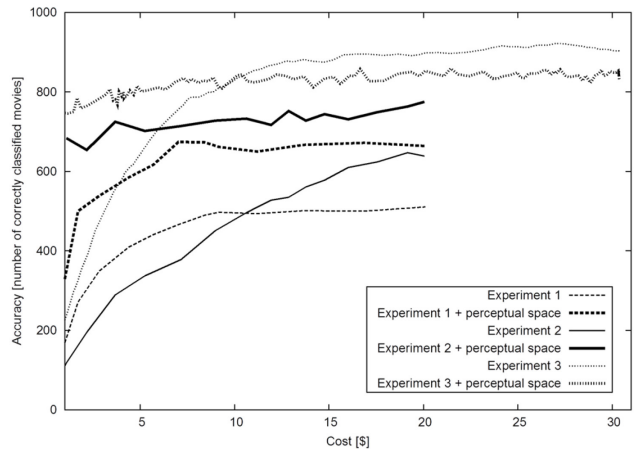

**Figure 4: Correctly classified movies over money spent**
Using perceptual spaces, a reliable classification is possible after for very low costs (i.e. after just few worker judgments.) Experiment 3 was more expensive due to higher workload.

When using all 893 movies classified in Experiment 1 for training (where only 533 of those have been correctly classified, resulting in an accuracy of 59.7%), using our perceptual space we can correctly classify 670 movies, where all 1,000 movies are classified. More interestingly, after only 15 minutes (or after spending $2.82 respectively), we can classify 538 movies correctly. By only using crowd-sourcing as in Experiment 1, only 349 movies had been correctly classified at that time.

These results allow many interesting observations and applications: In this experiment, the training set quality is very bad. Usually, reliable training sets with accuracy close to 100% are desirable. But even with this handicap, we show that we achieve significantly higher data quality than when just using crowd-sourcing alone. Also, we reach full 100% classification coverage, including even very obscure movies. Furthermore, after just $2.82, we can classify more movies correctly than when using $20 for pure crowd-sourcing (538 vs. 533). This opens up interesting options for cheaply expanding schemas even over large databases (see Section 4.3), and also leads to promising methods for improving quality of already available data (see Section 4.4).

*Experiment 5)*: Similarly to the previous experiment, we now boost the judgments of the more honest workers of Experiment 2. That is, compared to Experiment 4, the accuracy of the training set is increased from 59.7% to 79.4%. As expected, this also significantly improves the effectiveness our perceptual space based schema expansion: After 15 minutes (and due to less average judgments per hour, after spending only $2.16), already 654 movies can be classified correctly (compared to 195 classified movies in Experiment 1, and 538 in Experiment 4). When we use all classified movies as training data (after 120 minutes and $20 spent), 766 movies can be classified correctly. Here, we can already see that perceptual space extraction indeed benefits from higher quality training data. The full extent of improvements using small high quality training sets is analyzed in Section 4.3.

*Experiment 6)*: We now use the judgments of Experiment 3 as training input. These have been crowd-sourced under rigid control (and by looking up the "correct" answer in the Web), which would normally not have been possible for other perceptual data. The resulting training set has a final accuracy of 93.5% with 966 movies. Using this training set, an interesting observation can be made: After 15 minutes (and due to the low speed of this evaluation after just $0.32), already 732 movie could be classified correctly (at that time, the training set contained only 82 movies). Thus, very early on, the classification quality was very high for low costs. However, when using all 966 movies for training (with 93.5% accuracy), the quality of perceptual space classification is lower than the crowd-sourced quality. Using the space, 831 movies could be classified correctly (83% accuracy; higher than the accuracy in the last two experiments, but lower than the training set). This observation can be attributed to the fact that our approach provides an approximation based on highly-compressed data encoded in the perceptual space. Highly accurate and complete (and thus also very expensive) crowd-sourcing tasks are therefore hard to outperform.

In summary, the strongest application scenarios for perceptual spaces lie in (a) performing schema expansions involving perceptual attributes for large databases at very low costs with good accuracy and (b) improving the result quality of perceptual data obtained by direct crowd-sourcing. If used for improving data quality, additional steps are required (see Section 4.4).

### 4.3 Automatic Schema Expansion from Small Samples

We have seen that a large number of missing attribute values can be efficiently extrapolated from small crowd-sourced training samples. Now, we perform a more thorough investigation of this effect in a controlled environment. In particular, we demonstrate the effectiveness of very small training samples for deriving attribute values from perceptual spaces for even large databases.

By cross-referencing the three movie databases we used to construct our ground truth, we have been able to identify six genres that are shared by all:[6] Comedy, Documentary, Drama, Family,

---

[6]As the genre classification systems differ significantly between the databases, we have not been able to identify more genres that show a clear one-to-one correspondence. Equating genres that are only similar

545

**Table 3. Automatic schema expansion from small samples**
g-mean measures for expanding 10,562 movies using $n$ positive and $n$ negative training examples.

| Genre | Random | Perceptual space | | | Metadata space | | | Reference | | |
|---|---|---|---|---|---|---|---|---|---|---|
| | | $n=10$ | $n=20$ | $n=40$ | $n=10$ | $n=20$ | $n=40$ | Netflix | RT | IMDb |
| Comedy | 0.50 | 0.58 | 0.70 | 0.76 | 0.46 | 0.30 | 0.28 | 0.85 | 0.97 | 0.96 |
| Documentary | 0.50 | 0.73 | 0.81 | 0.84 | 0.64 | 0.63 | 0.62 | 0.96 | 0.99 | 0.97 |
| Drama | 0.50 | 0.60 | 0.66 | 0.73 | 0.51 | 0.45 | 0.49 | 0.86 | 0.90 | 0.92 |
| Family | 0.50 | 0.82 | 0.86 | 0.88 | 0.44 | 0.43 | 0.43 | 0.95 | 0.97 | 0.95 |
| Horror | 0.50 | 0.83 | 0.86 | 0.87 | 0.53 | 0.32 | 0.43 | 0.92 | 0.98 | 0.97 |
| Romance | 0.50 | 0.56 | 0.68 | 0.73 | 0.45 | 0.35 | 0.38 | 0.91 | 0.83 | 0.92 |
| Mean | 0.50 | **0.69** | **0.76** | **0.80** | 0.50 | 0.41 | 0.44 | 0.91 | 0.94 | 0.95 |

Horror, and Romance. For each genre, we created a ground truth classification by taking the majority vote over all three databases.

For a fully controlled experiment of the accuracy of automatically derived genre assignments, this time we abstained from experimenting on crowd-sourcing platforms. Instead, we created varying training sets directly from the reference data (thus showing 100% accuracy): For each genre, we randomly picked $n \in \{10, 20, 40\}$ positive and $n$ negative training examples from the full data set of 10,562 movies. For real word applications, such small and high quality samples can be provided directly by the crowd by careful selection of workers and elaborate quality control. Based on the representation of movies in the perceptional space, we then used a SVM classifier (see Section 4.2) to assign binary genre labels to all remaining movies and compared them to our ground truth.

For rigorously measuring the classification accuracy in this controlled experiment, we did not just use a simple accuracy measue as we did previously (i.e. relative number of correct classifications). This is because there is a substantial imbalance between positive and negative genre assignments in the data set, i.e. some genres are significantly more common than others. Using simple accuracy would result in the strange situation that a naïve classifier assigning the label *not Horror* to every movie would achieve an astonishing accuracy of 90%, as only 10% of all our movies are horror movies. A popular measure of classification performance in the presence of class imbalance is the g-mean measure [20], which is the geometric mean of sensitivity (accuracy on all movies truly belonging to the genre) and specificity (accuracy on all movies truly not belonging to the genre), As the g-mean punishes significant differences between sensitivity and specificity, the above naïve classifier would achieve 0% g-mean. Similarly, a random baseline (assigning each label with 50% probability) would achieve a g-mean of 50%.

In the following experiment we show that the high quality of our approach can indeed be attributed to the semantics of the perceptual space, and that similar quality cannot be achieved by simply relying on already available meta-data. Therefore, we compare the perceptional space to the information space spanned by ordinary movie metadata. This is implemented by using Latent Semantic Indexing (LSI) [21] to generate a 100-dimensional "metadata space" from movie attributes like title, plot, main actors, directors, year, runtime, and country as recorded in IMDb. Intuitively, this space captures the essence of the stored attributes in the database. Finally, we train an additional SVM classifier with the same training data as before, but this time using the metadata space.

(e.g., "Action" in IMDb and "Action & Adventure" in RT) would compromise our ground truth.

Table 3 shows our results for all experiments different genres and different choices of $n$. Each entry represents the mean of over 20 random repetitions in order to cancel out the influence of a single training set. Each computation took about 3 seconds on a typical notebook computer. Furthermore, we also include the g-mean accuracy of the individual expert databases IMDb, Netflix, and RT with respect to our reference data. Although the reference was created using these three datasets, they do not always agree on a genre classification. Thus, even when using these sources individually, g-mean values between only 0.91 and 0.95 with respect to the majority can be achieved. Furthermore, we can expect that any additional data source not used to create the reference, but also relies on highly paid experts and manual curation, will likely show g-mean values below 0.9.

But when using perceptual spaces, an astonishing g-mean of 0.8 could be achieved for $n = 40$ (i.e. when using 80 movies in the training sample). This means that just 0.75% of all items need to be judged manually by the crowd, all other items can be expanded automatically with high quality. This is also significantly less than the 10% recommended data for creating gold questions in regular crowd-sourcing tasks as shown in Experiment 3.

Furthermore, we can show that approaches based on classification using metadata and LSI lead to surprisingly bad results (g-mean between 0.41 and 0.50), and show even worse accuracy than randomly applying labels. This is a consequence of severe overfitting to the training data, also leading to a large variance of g-means across different random samples (standard deviation of about 0.20), while the accuracy using the perceptual space almost stays the same (standard deviations of about 0.02). This is due to the SVM trying to mine information out of the metadata space that is just not there—high-level perceptual judgments like genre classifications can only be given by humans who actually watched the movie and are not contained in the factual metadata.

To conclude, we found that schema expansion tasks on perceptual attributes can be performed at low cost and high speed by effectively leveraging the power of the crowd. The key to success are perceptual spaces, which can easily be built from the Social Web. The data quality achieved is close to human expert performance.

### 4.4 Automatic Identification of Questionable HIT Responses

Our experiments in Section 4.1 clearly showed that assuring data quality is a severe problem in crowdsourcing tasks on perceptual database attributes. This problem is caused by malicious crowd-workers, but also by users who are not domain experts. In any case, no matter what the source, we need a method that automatically identifies questionable HIT responses, which then can be sent to other crowd-workers for verification.

Table 4. Automatic identification of questionable HIT responses
Each entry is a precision/recall pair.

| Genre | Perceptual space | | | Metadata space | | |
|---|---|---|---|---|---|---|
| | $x = 5\%$ | $x = 10\%$ | $x = 20\%$ | $x = 5\%$ | $x = 10\%$ | $x = 20\%$ |
| Comedy | 0.30 / 0.83 | 0.47 / 0.83 | 0.66 / 0.82 | 0.07 / 0.53 | 0.09 / 0.43 | 0.15 / 0.39 |
| Documentary | 0.58 / 0.96 | 0.72 / 0.96 | 0.81 / 0.94 | 0.27 / 0.59 | 0.19 / 0.32 | 0.17 / 0.23 |
| Drama | 0.21 / 0.77 | 0.36 / 0.77 | 0.55 / 0.77 | 0.05 / 0.51 | 0.10 / 0.51 | 0.21 / 0.51 |
| Family | 0.73 / 0.95 | 0.83 / 0.96 | 0.84 / 0.95 | 0.00 / 0.07 | 0.01 / 0.08 | 0.17 / 0.24 |
| Horror | 0.62 / 0.93 | 0.77 / 0.93 | 0.84 / 0.93 | 0.01 / 0.10 | 0.06 / 0.18 | 0.09 / 0.18 |
| Romance | 0.29 / 0.88 | 0.46 / 0.88 | 0.65 / 0.87 | 0.14 / 0.57 | 0.13 / 0.36 | 0.15 / 0.29 |
| **Mean** | **0.46 / 0.88** | **0.60 / 0.89** | **0.73 / 0.88** | 0.09 / 0.40 | 0.10 / 0.31 | 0.16 / 0.31 |

In the following, we demonstrate how questionable HIT responses can reliably be identified by comparing the actual responses to the structure of the perceptual space. To ensure a controlled environment, we again rely on the six genres and the corresponding ground truth data sets as used in the previous section. For each genre, we assume that we are given a largely truthful classification of movies performed by the crowd, which contains a fixed number of misclassifications to be identified (e.g., 20% of all movies). We create such data sets by taking our ground truth derived from the three expert databases and swapping the classification labels of a randomly chosen set of $x\%$ of all 10,562 movies. Consequently, $(100 - x)\%$ of all labels are correct.

Given such a data set, to identify all incorrectly labeled movies, we trained an SVM classifier on the perceptual space representation of all 10,562 movies. Then, we marked those movies whose label in the given data set differs from the SVM model's prediction, i.e., we identify all movies whose crowd label contradicts the movie's position in the perceptual space. For example, a movie labeled as "Action" by the crowd but surrounded by non-Action movies in the perceptual space most likely is not an Action movie.

We measured the quality of our results by computing the precision and recall with respect to the ground truth classification. As in the previous section, we compared the perceptual space to a "metadata space" derived from the detailed explicit description of each movie. Also, we investigated the performance of our method on different degrees of data quality (modeled by the parameter $x$). Here, we chose $x = 5$, $x = 10$, and $x = 20$.

Table 4 shows the results of our experiments. Each entry is the mean over 20 independent runs (i.e., 20 different sets of randomly swapped labels). Clearly, using perceptual spaces, even with 20% wrong judgments, 88% of the swapped labels can reliably be detected. Here, the precision is 73%, i.e., just a few labels have been incorrectly flagged as being incorrect. Therefore, data quality can be improved cheaply by crowdsourcing additional judgments for all tuples flagged by our method. Also, the metadata space again proves to be prone to overfitting and thus vastly inferior to the perceptual space.

We conclude that perceptual spaces prove to be highly beneficial when questionable responses need to be identified in HIT tasks. By reevaluating those responses in a new crowd task, data quality can be increased significantly in crowd databases. At the same time, by focusing on questionable responses only, this increase of quality comes with minimal costs.

## 4.5 Exploring Other Domains

In the recent years, almost all modern websites or e-commerce platforms dealing with consumer goods adopted at least basic community features. In its simplest and easiest-to-use form, users can rate individual items. This information is typically used for deriving an aggregated community vote for each item. Surprisingly, most sites diligently collect vast amounts of rating data, but do use this data for further applications yet. Therefore, most of the data remains hidden in the databases of the platform providers (in these cases, only the providers themselves could derive perceptual spaces from the data), but sometimes, all individual user-rating pairs have been made publicly available. Some well-known examples are the iTunes Store (ratings of music, videos, and mobile apps), amazon.com (ratings of almost anything), Google Places, yelp.com (ratings of locations such as restaurants, bars, and shops), and ign.com (ratings of computer games).

To test whether our approach can be applied effectively to other domains, we crawled two vastly different data sets. The first one consists of San Francisco restaurant ratings from yelp.com (3,811 restaurants; 128,486 users; 626,038 ratings). The second one consists of board game ratings from boardgamesgeek.com (32,337 games; 73,705 users; 3,536,455 ratings). In both services human editors manually created detailed descriptions of all items, which we used for measuring the performance of our method. We chose 10 binary categories for restaurants and 20 for board games.

We repeated the experiment "automatic schema extension from small samples" (Section 4.3) on the new data. The measured g-mean accuracies are slightly lower than those on the movie domain. However, we did not invest much time in tuning model parameters or the SVM classifier. Also, for training the classifier, in each domain we had to rely on the (possibly inaccurate) catego-rization from a single website and did not pool the opinions of several independent metadata providers as we did for movies. Details for some representative categories are given in Table 5 and Table 6. Again, we observe that truly perceptual categories such as "party game" can be identified much better than purely factual ones such as "modular board."

Table 5. Results for restaurants
g-mean measures using $n$ positive and $n$ negative training examples.

| | Perceptual space | | |
|---|---|---|---|
| Category | $n = 10$ | $n = 20$ | $n = 40$ |
| Ambience: Trendy | 0.62 | 0.64 | 0.68 |
| Attire: Dressy | 0.60 | 0.68 | 0.70 |
| Category: Fast Food | 0.67 | 0.75 | 0.82 |
| Good For Kids | 0.55 | 0.62 | 0.66 |
| Noise Level: Very Loud | 0.62 | 0.71 | 0.79 |
| … | … | … | … |
| Mean | 0.62 | 0.67 | 0.75 |

Table 6. Results for board games
g-mean measures using $n$ positive and $n$ negative training examples.

| | Perceptual space | | |
|---|---|---|---|
| Category | $n = 10$ | $n = 20$ | $n = 40$ |
| Collectible Components | 0.56 | 0.67 | 0.72 |
| Children's Game | 0.61 | 0.64 | 0.71 |
| Party Game | 0.63 | 0.67 | 0.71 |
| Modular Board | 0.47 | 0.48 | 0.52 |
| Route/Network Building | 0.72 | 0.75 | 0.77 |
| Worker Placement | 0.72 | 0.78 | 0.80 |
| … | … | … | … |
| Mean | 0.63 | 0.68 | 0.73 |

We conclude that our approach indeed generalizes to domains beyond movies, but might need moderate tuning effort.

## 5. OPEN ISSUES

In this section, we discuss additional issues that may arise when applying our approach to different data sets and offer some solutions. We will also present and discuss possible extensions and modifications of our approach.

**Scarce data.** For less popular domains, the potentially small number of users and ratings can be problematic. Clearly, only little can be learned about an item's properties using perceptual spaces if no or only very few ratings are available. However, the previous section already indicates that even for niche domains such as board games, data sets with a sufficient number of ratings can be found. Still, even when there really are only small numbers of users and ratings, recent work in recommender systems shows that much can be learnt from the available data as long as there are some active core users, who rated a large number of items each [22]. Thus, only in scenarios with highly inactive and passive communities our approach will not be able to bring up the desired results.

**Advanced perceptual spaces.** To illustrate the general applicability of our approach, we built on a simple but well-established model for deriving our perceptual spaces. Of course, there is room for improvement by incorporating additional features into the model. For example, users could be represented by more than a single point in space to model diverse interests [23]. Also, changing taste over time could be integrated by respecting the time a rating was given [24]. Finally, additional meta-data on items beyond plain ratings could be incorporated [25]. However, these issues are beyond the scope of this paper.

**Semi-supervised learning.** In most settings, the database contents are likely to change at a high frequency. Therefore, after building the perceptual space, we usually know for what items missing perceptual attributes need to be determined. This makes our setting very similar to semi-supervised learning tasks, where in addition to the labeled data a large amount of unlabeled data is used for training the classifier [26].A highly effective and very popular semi-supervised learning technique are transductive support vector machines (TSVMs) [27]. As TSVMs directly extend the supervised SVM classifier we used in our experiments, we repeated the experiment on schema extension reported in Section 4.3 with the TSVM classifier provided by the well-known SVMlight library[7]. We observed almost the same classification accuracies as those reported in Table 3 (average g-means of 0.70, 0.77, and 0.79, respectively), but were surprised by the huge differences in runtime. While the original SVM (also from the SVMlight library) required about 3 seconds for each classification task, the TSVM usually needed around 90 minutes. This huge difference can be attributed to the large variations in input size: only a few training examples with the SVM, but the entire database with the TSVM (quadratic complexity). As an alternative, we also tried the reference implementation of the more advanced CCCP-SVM approach, which has been reported to work significantly faster [28]. However, in our setting we still observed runtimes of about 20 minutes per classification task, which is definitely too slow for augmenting a real-time crowdsourcing task. We conclude that the application of semi-supervised learning techniques is tempting but requires a careful choice of methods to avoid scalability problems.

## 6. RELATED WORK

The benefits of crowd-sourcing have been demonstrated to solve a variety of problems in many different disciplines, e.g., to assess the usability of user interfaces [29] or to quickly gather information required for disaster response [30]. Very recently, the crowd-enabled databases [1–3] pointed out the huge potential of crowd-sourcing for database applications.

The issue of crowd data quality has been recognized and addressed in several papers. For example, in [31] it is shown that the selective repetition of some HITs may significantly improve data quality when dealing with non-expert workers. In [32] and [33] different strategies are developed to infer a single reliable judgment from conflicting responses to the same HIT, mostly by extending the majority voting scheme. Finally, [34] investigates the effect of financial incentive and arrives at a surprising conclusion: Increased payment increases the quantity, but not the quality of HITs. We are not aware of any work that exploits the Social Web to tackle the problem of data quality in crowd-sourcing.

Perceptual spaces are rooted in theories of human pattern recognition and categorization and are a popular model in cognitive psychology [35]. However, the perceptual spaces used in psychology are typically created from extensive pen-and-paper surveys and thus tend to be extremely small. Recently, some algorithms have been developed in the area of recommender systems that derive personalized item rankings from rating data and create perceptual spaces as a by-product [10]. However, only a small number of papers yet investigated the inner structure of these spaces and their use for applications beyond pure recommendation. Most work in this direction focuses on comparing spaces created by different methods for means of classification [11], [36].

## 7. CONCLUSION

Recently, crowd-enabled databases have been introduced as powerful and innovative extensions to relational databases that can provide cognitive operators and can also complete missing values and tuples at query time. In this paper, we extended this concept even further by also allowing the expansion of database schemas with additional attributes in a query-driven fashion. We have shown that this challenging feat cannot be achieved by naïve crowd-sourcing alone when considering performance and quality constraints. Especially when dealing with perceptual properties, which rely on honest judgments, serious problems may arise with respect to data quality. Additionally, due to limited expert knowledge of workers, it is hard to obtain reliable judgments for lesser-known items. This limits scalability even further.

Therefore, we propose to tap into the vast wealth of human-generated data available in the Social Web. From this data, perceptual spaces can be built, which contain human judgments of characteristics of a large number of items in distilled form. By using the example of ubiquitous rating data, we demonstrated how to build such a perceptual space for the movie domain. By training a classifier via crowdsourcing, we could successfully show that our approach can reliably expand schemas of even large databases with high reliability and at very low costs. Furthermore, we demonstrated how this method can be applied to approach general open issues of crowd-enabled databases like data quality.

---

[7] http://svmlight.joachims.org/

We plan to extend our work by integrating some quality assurance methods (cf. Section 5). Moreover, we are currently investigating how perceptual spaces can be built efficiently from other data sources such as textual reviews and recommended links.